\documentclass[prb,showpacs,superscriptaddress,floatfix,twocolumn]{revtex4}
\usepackage{amsfonts}
\usepackage{stmaryrd}
\usepackage{bbm}
\usepackage{mathrsfs}
\usepackage{tipa}
\usepackage{amssymb}
\usepackage{txfonts}
\usepackage{graphicx}
\usepackage{dcolumn}
\usepackage{epstopdf}
\usepackage[colorlinks,linkcolor=blue,urlcolor=blue,citecolor=blue]{hyperref}
\newcommand{\PreserveBackslash}[1]{\let\temp=\\#1\let\\=\temp}
\newcolumntype{C}[1]{>{\PreserveBackslash\centering}p{#1}}
\newcolumntype{R}[1]{>{\PreserveBackslash\raggedleft}p{#1}}
\newcolumntype{L}[1]{>{\PreserveBackslash\raggedright}p{#1}}

\begin{document}

\newcommand*{\cm}{cm$^{-1}$\,}


\title{Unconventional charge-density wave in Sr$_{3}$Ir$_{4}$Sn$_{13}$ cubic superconductor revealed by optical spectroscopy study}

\author{A. F. Fang}
\author{X. B. Wang}
\author{P. Zheng}
\affiliation{Beijing National Laboratory for Condensed Matter
Physics, Institute of Physics, Chinese Academy of Sciences,
Beijing 100190, People's Republic of China}

\author{N. L. Wang}
\affiliation{Beijing National Laboratory for Condensed Matter
Physics, Institute of Physics, Chinese Academy of Sciences,
Beijing 100190, People's Republic of China}
\affiliation{Collaborative Innovation Center of Quantum Matter, Beijing, China}


\begin{abstract}
Sr$_{3}$Ir$_{4}$Sn$_{13}$ is an interesting compound showing a coexistence of structural phase transition and superconductivity. The structural phase transition at 147 K leads to the formation of a superlattice. We performed optical spectroscopy measurements across the structural phase transition on single crystal sample of Sr$_{3}$Ir$_{4}$Sn$_{13}$. The optical spectroscopy study reveals an unusual temperature-induced spectral weight transfer over broad energy scale, yielding evidence for the presence of electron correlation effect. Below the structural phase transition temperature an energy gap-like suppression in optical conductivity was observed, leading to the removal of partial itinerant carriers near Fermi level. Unexpectedly, the suppression appears at much higher energy scale than that expected for a usual charge density wave phase transition.
\end{abstract}

\pacs{78.20.-e, 71.45.Lr, 74.70.-b}

\maketitle
Transition metal compounds are well known for exhibiting rich and interesting physical properties such as charge or spin density wave\cite{charge-1,charge-2,spin-1,spin-2}, superconductivity\cite{sc-1,sc-2} and modulated lattice distortion \cite{superlattices-1,superlattices-2}. The interplay between two very different electronic phenomena is one of the fundamental interests in condensed-matter physics. For example, the layered transition metal dichalcogenides TX$_2$ (T: transition metal, X: chalcogen) ordinarily possessing the charge-density wave (CDW) instability and its coexistence/competition with superconductivity are central characteristics of this family.\cite{interplay-1,interplay-2,interplay-3} Lately, the 5d transitional metal ditelleride Ir$_{1-x}$Pt$_x$Te$_2$ attracted particular attention for the competing phenomenon between structural instability and superconductivity.\cite{superlattices-1,IrTe2-1,IrTe2}

A very recent example is the cubic superconductor Sr$_{3}$Ir$_{4}$Sn$_{13}$ with $T_c=$ 5 K.\cite{growth2,HC,SrIrSn-pressure} This compound undergoes a structural phase transition at $T^*\simeq$ 147 K, evolving from the $I$ structure ($Pm\overline{3}n$) to a $I'$ structure ($I\overline{4}3d$) accompanying with a doubling of the unit cell on decreasing temperature across the transition.\cite{SrIrSn-pressure} Concerning the structural phase transition, Klintberg $et$ $al.$ found that a superlattice distortion emerged below the structural phase transition and assigned to a CDW modulation.\cite{SrIrSn-pressure}  A muon spin rotation ($\mu$SR) study on isovalent compound Ca$_{3}$Ir$_{4}$Sn$_{13}$ ($T_c=$ 7 K, $T^*\simeq$ 33 K) confirmed that the structural phase transition was a nonmagnetic transition,\cite{usr} making the charge-density wave scenario more plausible and excluding the possibility of ferromagnetic spin-fluctuation scenario.\cite{Ferro} Despite the extensive studies on (Sr,Ca)$_{3}$Ir$_{4}$Sn$_{13}$, the origin of the structural transition and its relation to superconductivity are yet to be clarified. On the other hand, it is well known that CDW often occurs in low dimensional electronic systems, while such instability is not expected to develop in a three-dimensional (3D) system. Sr$_{3}$Ir$_{4}$Sn$_{13}$ supplies us a precious opportunity to investigate the interplay between possible CDW instability and superconductivity in a 3D cubic compound. It is also argued that the structural phase transition in (Sr,Ca)$_{3}$Ir$_{4}$Sn$_{13}$ is a pressure-induced structural quantum critical point,\cite{SrIrSn-pressure} which makes the compounds more interesting.

It is of great significance to understand the origin of the structural phase transition, since it is the essential step towards understanding the mechanism of superconductivity. In this work, we report an optical spectroscopy investigation on well characterized Sr$_{3}$Ir$_{4}$Sn$_{13}$ single crystal samples. We show that the structural phase transition is associated with the formation of partial energy gap at the Fermi surface. A significant number of itinerant carriers are lost due to the opening of the gap. We suggest an unconventional CDW scenario to account for the  second order structural phase transition.

\begin{figure}
\includegraphics[clip,width=2.5in]{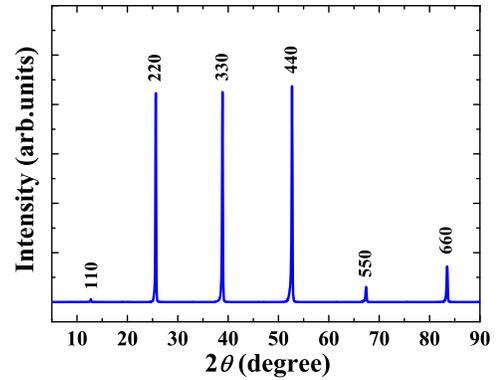}
\caption{(Color online)
X-ray diffraction patterns of Sr$_{3}$Ir$_{4}$Sn$_{13}$ single crystals at room temperature.}
\end{figure}

\begin{figure}
\includegraphics[clip,width=3.45in]{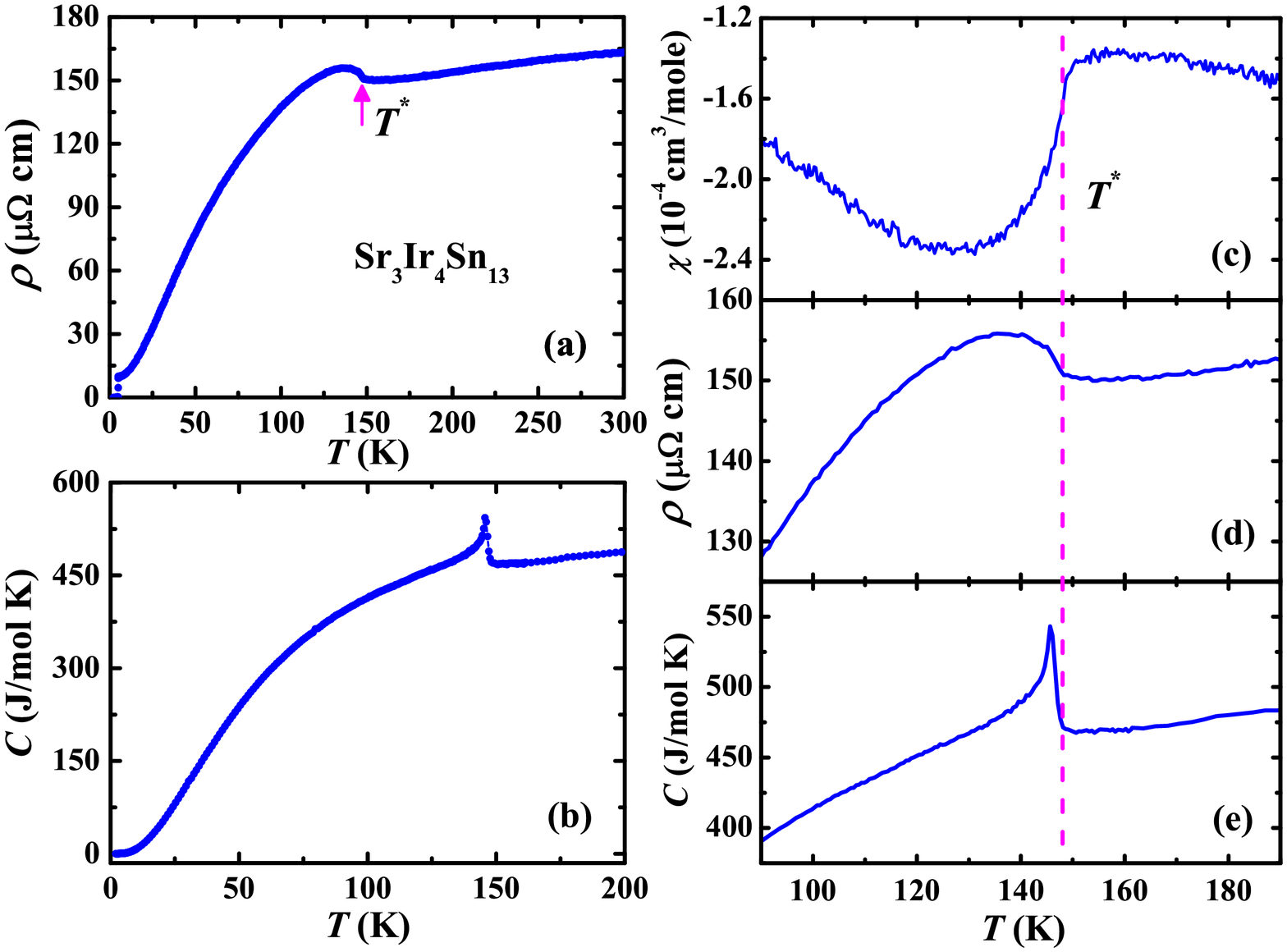}
\caption{(Color online)
Physical properties of Sr$_{3}$Ir$_{4}$Sn$_{13}$ single crystals. (a) Resistivity $\rho$ versus temperature in the range of 2 K to 300 K. (b) Temperature dependence of specific heat \emph{C} with zero applied field. (c)(d)(e) The temperature dependent magnetic susceptibility, resistivity and specific heat from 90 K to 190 K, respectively. The dash line indicates the phase transition begins at $T^*$ = 147 K. }
\end{figure}

Sr$_{3}$Ir$_{4}$Sn$_{13}$ single crystals were synthesized by self-flux technique in a procedure similar to earlier reports.\cite{growth1, growth2} The Sn flux  was initially removed from single crystals using a centrifuge at approximate 575 $^\circ \mathrm{C}$, and then the residual flux was etched in concentrated hydrochloric acid. Shiny 3D samples were successfully extracted with maximum dimensions of 3 $\times$ 5 $\times$ 5 mm$^3$. In order to keep the whole sample at the same temperature with the sample holder during optical spectroscopy measurements, we picked a single crystal with a large flat surface and polished its opposite plane until a thin sheet obtained. The flat surface was identified to be (110) plane by single crystal X-ray diffraction (XRD) measurements, as displayed in Figure 1. The \emph{I} phase structure with space group $Pm\bar{3}n$ and lattice parameter a=9.8156 ${\AA}$ were obtained from single crystal XRD data at room temperature.

The dc resistivity measurements were conducted on a commercial Quantum Design physical properties measurement system (PPMS) by a four-probe method. The specific-heat data were obtained by a relaxation-time method using PPMS. The magnetic susceptibility was measured on a quantum design superconducting quantum interference device vibrating sample magnetometer system (SQUID-VSM). Figure 2 (a) presents the temperature dependence
of resistivity $\rho$ for Sr$_{3}$Ir$_{4}$Sn$_{13}$ single crystals. A distinct anomaly is observed at $T^*$ = 147 K on cooling, while no obvious hysteresis of $T^*$ is found upon warming (not shown). The specific heat $\emph{C}_p$ also reveals a sharp anomaly at corresponding $T^*$, as shown in Figure 2 (b). The characteristic $\lambda$-like shape of the sharp anomaly identifies the $T^*$ anomaly as a second order phase transition. Figure 2 (c) shows that the $T^*$ phase transition leads to a sudden drop in magnetic susceptibility, which is similar with that of IrTe$_2$.\cite{IrTe2} As exhibited in Figure 2 (c)(d)(e), the phase transition creates clear anomalies at the same $T^*$ for magnetic susceptibility, resistivity and specific heat.

As reported in the previous work, the structural phase transition at 147 K results in a $1.1\times10^{-4}$ emu/mol change of Pauli susceptibility.\cite{SrIrSn-pressure} Our result is consistent with it. Another striking point is that the magnetic susceptibility remains negative even above the phase transition. As is well known, the closed electronic shells or fully occupied bands would contribute to the diamagnetism, so called Larmor diamagnetism. Since Iridium is a 5$d$ transition metal, additionally, and the crystals contain Sn$_{12}$ cages in the crystal structure, Larmor diamagnetism is probably sufficient to exceed the paramagnetic Pauli spin susceptibility $\chi_P$ of the conduction carriers. Therefore, the magnetic susceptibilities of Sr$_{3}$Ir$_{4}$Sn$_{13}$ compounds are diamagnetic. Considering the result (obtained below) that about 30 \% itinerant carriers at Fermi surfaces are lost and $\chi_P$ is proportional to the conduction carriers, the diamagnetism is further enhanced below the phase transition.

Another significant phenomenon revealed by electrical resistivity $\rho(\emph{T})$ is a superconducting phase transition at $T_c$ = 5 K.
Figure 3 (a) shows the temperature dependent low temperature magnetic susceptibility for Sr$_{3}$Ir$_{4}$Sn$_{13}$ single crystals in H = 50 Oe
parallel to the (110) plane with zero field- cooled (ZFC) and field-cooled (FC) processes. Low temperature magnetic susceptibility suggests Meissner effect of the superconductivity below 5 K. The obvious peak of specific heat at $T_c$ shown in the Figure 3 (b) confirms the bulk superconductivity for Sr$_{3}$Ir$_{4}$Sn$_{13}$ single crystals. It is known that the specific heat at low temperature could be approximately  written by $C_p/T = \gamma+\beta T^2+\eta T^4$, with the three terms of the right side representing electron, phonon and the anharmonic phonon contributions respectively. The linear fit of normal state C/T as a function of $T^2$ [solid line in Fig. 3 (b)] provides the electronic specific heat coefficient $\gamma$ = 40.0(1) mJK$^{-2}$mol$^{-1}$ , giving a good agreement with previous measurements.\cite{HC} Since $\gamma$ is proportional to $N(E{^0_F})$ (the electronic density of states near the fermi level), $N(E{^0_F})$ should be rather large, indicating presence of relatively narrow energy band. Therefore, electron correlation effects might not  be weak in Sr$_{3}$Ir$_{4}$Sn$_{13}$.

\begin{figure}
\includegraphics[clip,width=2.4in]{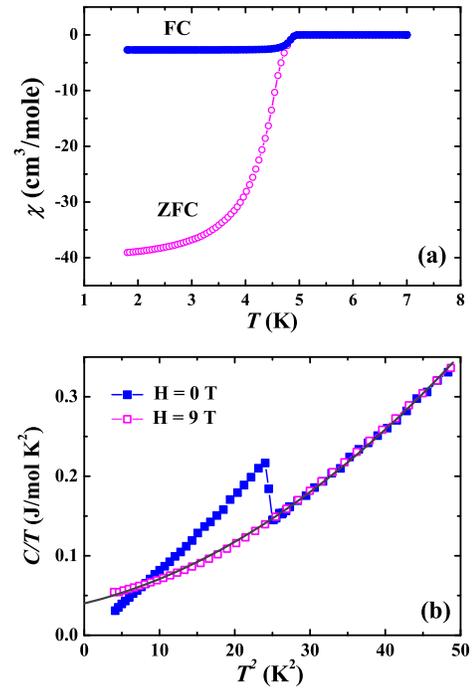}
\caption{(Color online)
(a) Low temperature magnetic susceptibility $\chi$ versus temperature  for Sr$_{3}$Ir$_{4}$Sn$_{13}$ single crystals in $H=50$ Oe parallel to the (110) plane with zero field-
cooled (ZFC) and field-cooled (FC) processes. (b) Specific heat of Sr$_{3}$Ir$_{4}$Sn$_{13}$ in the plot of $C/T$ versus $T^2$ with
0 T and 9 T applied field perpendicular to the (110) plane; the solid line is the fitting result using the relation $C=\gamma T + \beta T^3 + \eta T^5$ after the superconductivity is completely
suppressed by a magnetic field of 9 T. }
\end{figure}

The temperature-dependent (110)-plane optical reflectance data were measured on Bruker 113v, Vertex 80v and a grating type
spectrometers on shiny flat surfaces of Sr$_{3}$Ir$_{4}$Sn$_{13}$ superconducting single
crystals over broad frequency range from 40 to 50000 \cm (\cm relates to 1/ $\lambda$ in this paper).
We firstly obtained the reflectivity $R(\omega)$ by an $in $ $situ$
gold and aluminum overcoating technique, further getting the real part of conductivity $\sigma_1(\omega)$ by the Kramers-Kronig transformation of $R(\omega)$. Figure 4 displays the reflectance and optical conductivity spectra for Sr$_{3}$Ir$_{4}$Sn$_{13}$ single crystals by systematically varying temperature from 10 K to 300 K. All the $R(\omega)$ approach to unity at zero frequency both above and below the $T^*$ phase transition. A good metallic feature is obviously revealed, corresponding to the dc electric resistivity. There is a compelling phenomenon in the $R(\omega)$ spectra below the $T^*$ phase transition, i.e. a prominent suppression feature below 4000 \cm (marked by an arrow), especially that at 10 K. Clarifying the suppression feature is crucial for elaborating the evolution of the electronic structure across the transition.

The suppression feature is clearly reflected in the optical conductivity spectra as well, shown in Figure 4 (b). Apparent Drude components could be observed at low frequency in all spectral curves, indicating a good metallic response. Above the $T^*$ phase transition, the low frequency Drude component shows ordinary narrowing resulting from the reduced scatterings with lowering temperature, while an unconventional spectral weight transfer is observed from the low-energy mid-infrared region to the high-energy interband transition region above 5000 \cm. Cooling the sample across the $T^*$ phase transition, a significant spectral weight suppression develops in the Drude components and the suppressed spectral weight is transferred to the peak feature (marked by an arrow) forming at about 4000 \cm.
Since the Drude component represents the contribution from conduction electrons, the suppression of the Drude components observed in our optical spectroscopy experiments suggests the removal of a part of Fermi surface below the phase transition. At variance with other compounds possessing the first order structural phase transitions e.g. IrTe$_2$\cite{IrTe2} or BaNi$_2$As$_2$\cite{BaNiAs}, where the optical conductivity spectra develop sudden
and dramatic changes over broad frequencies across the phase
transitions owing to the reconstruction of the band structures, the present case is more similar to some CDW materials such as 2H-TaS$_2$\cite{TaS2} whose spectral suppression features are rather weak and evolves continuously with temperature. On the other hand, the unconventional spectral weight transfer also exists below the $T^*$ phase transition. This unconventional spectral weight transfer may be ascribed to the correlation effect, e.g. the intersite coulomb repulsion, similar with some charge-ordering systems.\cite{LaSrFeO,LaSrNiO,LaSrMnO}

\begin{figure}
\includegraphics[clip,width=3.3in]{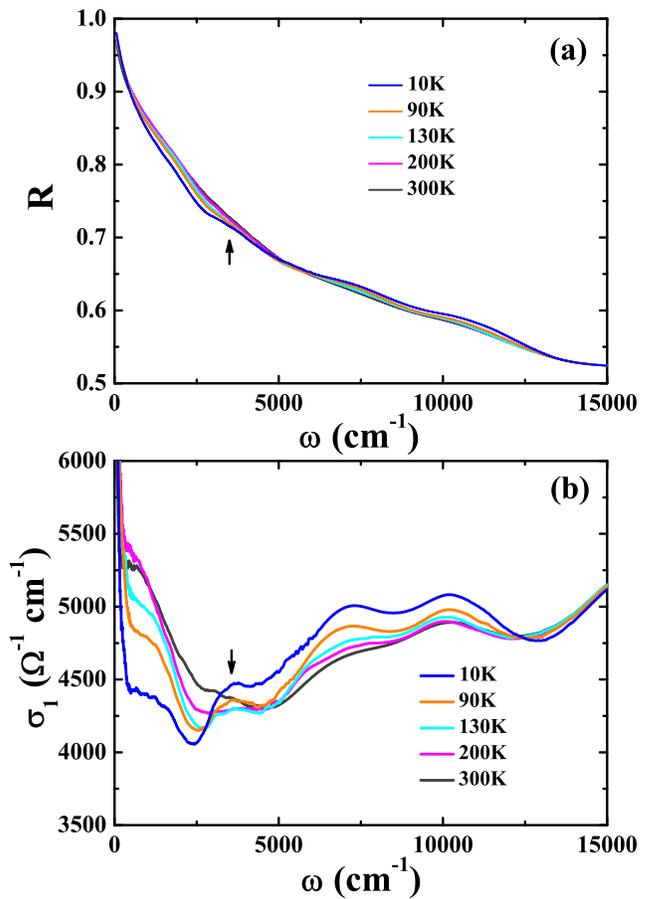}
\caption{(Color online)
(a) The (110)-plane reflectance spectra of Sr$_{3}$Ir$_{4}$Sn$_{13}$ single crystals at various temperatures with the black arrow marking the onset of suppression feature. (b) Optical conductivity $\sigma_1(\omega)$ up to 15000 \cm; the black arrow points at the newly developed peaks. }
\end{figure}

\begin{figure}
\includegraphics[clip,width=3.4in]{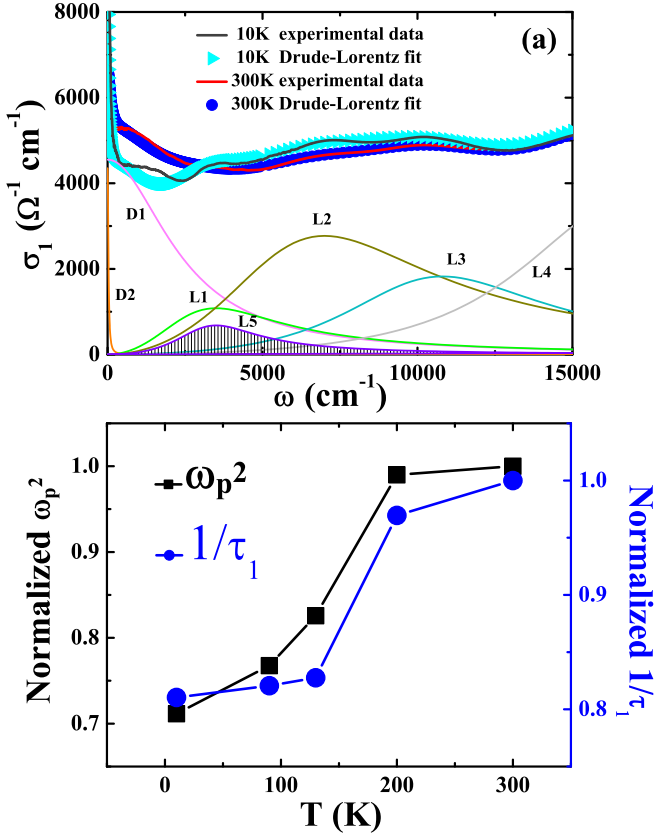}
\caption{(Color online)
(a) The experimental $\sigma_1(\omega)$ data at 10 K and 300 K up to 15000 \cm. The Drude (D1, D2) and the Lorentz (L1, L2, L3, L4, L5) terms acquired from a Drude-Lorentz fit for T=10 K below 15000 \cm are illustrated at the bottom. (b)Normalized total $\omega_p^2$ (black line) and $1/\tau_1$ (blue line). Both parameters are normalized to the
values of 300 K
 .}
\end{figure}

Aiming to further investigate the evolution of Drude components, the Drude-Lorentz model was applied to decompose the optical conductivity spectra into different
components. The dielectric function can be written as:\cite{densitywave-2}
\begin{equation}
\epsilon(\omega)=\epsilon_\infty-\sum_{i}{{\omega_{p,i}^2}\over{\omega_i^2+i\omega/\tau_i}}+\sum_{j}{{\Omega_j^2}\over{\omega_j^2-\omega^2-i\omega/\tau_j}},
\label{chik}
\end{equation}
where $\epsilon_\infty$ is the dielectric constant at high energy,  and the middle and last terms are the Drude and Lorentz components, respectively. The Drude components describe the contribution from conduction electrons, while the Lorentz components represent the interband transitions. As indicated by the band structure calculations, for Sr$_{3}$Ir$_{4}$Sn$_{13}$, several different bands cross the Fermi level and the Fermi surfaces are rather complicated.\cite{SrIrSn-pressure,SrIrSn-band} As a consequence, the conduction electrons are contributed by different bands and exhibit different behaviors. Therefore, we applied two Drude components analysis here. In order to reproduce the optical conductivity below 15000 \cm at 300 K, two Drude components, a narrow one and a broad one, and four Lorentz terms have to be used. The narrow one only occupies a small fraction of the spectral weight of the conduction electrons, while the broad Drude component takes most part of this spectral weight. However, an additional Lorentz component centered at 3500 \cm (L5 in Figure 5) should be added at temperatures below the phase transition. Actually, such two Drude components analysis has been widely applied to the multiband systems, for example Fe-based superconductors (although the Fe-pnictide superconductors usually have five Fermi surfaces (three hole type Fermi surfaces, two electron type Fermi surfaces)).\cite{prl-2008-1,twodrude,twodrude-1}

Figure 5 (a) illustrates the conductivity spectra at 10 K and 300 K together with the Drude-Lorentz fitting components for Sr$_{3}$Ir$_{4}$Sn$_{13}$. It is found that the two Drude components narrow with decreasing temperature because of the metallic response, while the gapping of the Fermi surfaces mainly occurs in the broad Drude component. The results indicate that the gapping of the Fermi surfaces caused by the T$^*$ phase transition mainly takes place on those Fermi surfaces where the electrons experience stronger scattering. Compared with the spectral weight distribution at 300 K, the spectral weight of D1 decreases and the suppression part of D1 is transferred to that of the added L5 as well as that of L2, L3 at 10 K. Considering that two Drude components contribute to the conductivity, the formula $\omega_p=\sqrt {{\omega_p{_1}^2} + {\omega_p{_2}^2}} $ could be used to obtain the overall plasma frequency $\omega_p$, and then $\omega_p \approx$ 30530 \cm at 300 K and $\omega_p \approx$ 25750 \cm at 10 K are estimated. Therefore, the ratio of the square of the plasma frequency at low temperature phase to that at 300 K is about 0.71, as shown in Figure 5 (b). It is well known that $ \omega _p^2 = 4\pi ne^2/m^*$, where n is the carrier density and $m^*$ is the effective mass. Provided that the effective mass of the itinerant carriers keeps unchanged, the optical measurements reveal that roughly 29\% itinerant carriers are lost obtained from the change of $\omega_p$  after the second order phase transition, which agrees with the previous electronic structure calculations.\cite{SrIrSn-pressure} However, the overall scattering rate cannot be estimated by a simple summation of the two scattering rates of two separate Drude components ($1/\tau=1/\tau_1+1/\tau_2$) that only works out for one Drude component with multiple scattering channels. For two Drude components analysis, the conductivity contributed by conduction electrons could be described by ${\sigma _D}{\rm{ = }}\frac{1}{{4\pi }}[\omega _{P1}^2/({\gamma _{D1}} - i\omega ) + \omega _{P2}^2/({\gamma _{D2}} - i\omega )]$, thus a single Drude component with re-weighted scattering cannot represent the two Drude Components. In addition, the narrow Drude component only shows ordinary narrowing with decreasing temperature due to the metallic response and the evolution of this scattering rate is not related with the $T^*$ phase transition. Therefore we just show the scattering rate $1/\tau_1$ of the broad Drude component in Figure 5 (b). The scattering rate $1/\tau_1$ is reduced by about 20 \% after the transition, indicating that the gap of partial Fermi surface might reduce the scattering channels.

Although the electronic structure calculations give an estimation of 30 \%
electronic density of states loss across the $T^*$ phase transition, there is no clue of the possible gap size at the Fermi surface supplied.\cite{SrIrSn-pressure} It is well known that the characteristic of a density wave order transition is the formation of an energy gap near the Fermi level, which leads to a pronounced peak just above the energy gap in $\sigma_1(\omega)$.\cite{densitywave,densitywave-1,densitywave-2} On this regard, the new peak formed below $T^*$ ($\sim$ 3500 \cm at 10 K) could be identified as CDW energy gap. However, the ratio of the energy gap relative to the transition temperature $2\triangle/k_BT_s$ reaches 33, which is significantly larger than the BCS mean field value of 3.5 for a density wave phase transition.\cite{densitywave} This implies that the $T^*$ phase transition temperature is much lower than the mean field transition temperature. Usually, this phenomenon is attributed to low-dimensional (e.g. 1D or 2D) electronic structure which gives rise to strong fluctuation effect and suppresses the actual ordering temperature. Apparently, this scenario could not be applied here, since Sr$_{3}$Ir$_{4}$Sn$_{13}$ compound is a 3D cubic material which is not supposed to cause strong fluctuation effect.

\begin{figure}
\includegraphics[clip,width=3.1in]{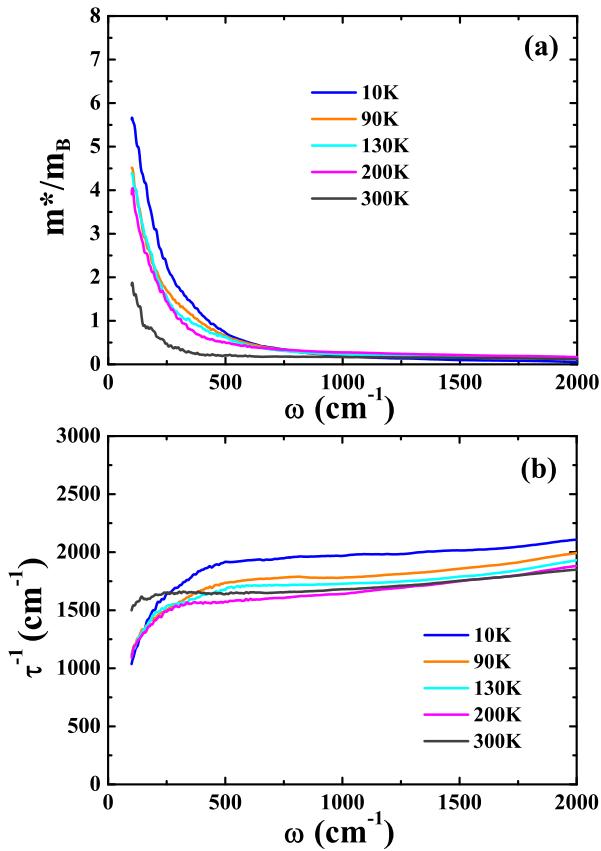}
\caption{(Color online)
The frequency dependence of the effective mass and (b) the scattering rate of Sr$_{3}$Ir$_{4}$Sn$_{13}$. }
\end{figure}

For correlated metals, one-component approach is also often used in the analysis of the infrared conductivity. In this
approach the simple Drude model is extended by making the damping term in the Drude formula
complex and frequency dependent.\cite{Timuskreview} This so-called extended Drude model has been
extensively applied to transition-metal compounds, heavy-fermion systems, high-Tc cuprates, and other systems with electrons
coupling strongly to certain bosonic excitations. Since the low temperature specific heat analysis reveals a
moderate enhancement of electronic specific heat coefficient in Sr$_{3}$Ir$_{4}$Sn$_{13}$,
we also conducted the extended Drude analysis of optical data and got frequency dependent scattering rate and the effective mass in terms of
conductivity spectra: \cite{Timuskreview}
\begin{equation}
\frac{1}{{\tau (\omega )}} = \frac{{\omega _p^2}}{{4\pi }}\frac{{{\sigma _1}(\omega )}}{{\sigma _1^2(\omega ) + \sigma _2^2(\omega )}},
\label{chik}
\end{equation}
\begin{equation}
\frac{{{m^*}}}{{{m_B}}} = \frac{{\omega _p^2}}{{4\pi \omega }}\frac{{{\sigma _2}(\omega )}}{{\sigma _1^2(\omega ) + \sigma _2^2(\omega )}},
\label{chik}
\end{equation}
where $\omega _p^2$ is the unscreened overall plasma frequency, m$_B$ the band mass. The obtained spectra of the effective mass and the scattering rate with $\omega _p$=25750 \cm (at 10 K) are displayed in Figure 6. Both the scattering rate and the effective mass show pronounced temperature dependence at low frequencies. Notably, the scattering rate is suppressed below about 300 \cm, and correspondingly the
effective mass is strongly enhanced. This energy is about one order smaller than the peak feature indicated in the above analysis. If this
energy scale is related to the CDW energy gap, it would provide a natural explanation for the value of $2\triangle/k_BT_s$ predicted by the BCS theory.
However, this is highly unlikely. In fact, the suppression feature is present clearly in the curves at 200 K and 300 K, which is much higher than the
phase transition temperature. From the conductivity spectra shown in Fig. 4 (b) we find clearly the presence of two Drude components. The strong
spectral change near 300 \cm in the effective mass and the scattering rate should be related to the two Drude components. On this basis, we believe that the Drude-Lorentz decomposition of $\sigma_1(\omega)$ is more appropriate here.

The surprisingly large value of energy gap is a very puzzling phenomenon. It points towards an unconventional driving mechanism for CDW phase transition in such material. It is worth noting that Varma and Simons proposed a strong-coupling theory of CDW transitions whose essential ingredients are the strong wave-vector dependence of electronically induced anharmonicity and mode-mode coupling.\cite{Varma} This microscopic theory permits an order of magnitude larger than the BCS mean field value of 3.5 and predicts phonons over a substantial part of the Brillouin zone are softened near the transition. The strong coupling picture is also supported by the specific heat data. From the electronic specific heat coefficient $\gamma=40 mJ/K^2 mol$ as derived from Fig. 3, we estimate electronic heat capacity at the CDW transition temperature roughly 6 $J/K mol$. The jump in the heat capacity at the transition observed from Fig. 2 is about $\Delta C\approx$60-70 $J/K mol$. This leads to the ratio of $\Delta C/C|_{T_{CDW}}\geq$10, which is much larger than the value (1.43) of simple mean field BCS theory for CDW phase transition. In addition, according to the Fermi-liquid theory, $\gamma$ is proportional to $n^{1/3}m^*$ ($\gamma = \frac{\pi^{2/3}k_B^2}{3^{2/3}\hbar^2}n^{1/3}m^*$) at low temperature. In combination with the renormalized plasma frequency ${\left( {\omega _p^*} \right)^2} = \frac{{120}}{\pi }\int_0^{{\omega _c}} {{\sigma _1}} (\omega )d\omega  = \frac{{4\pi n{e^2}}}{{{m^*}}}$, we can obtain the values of carrier density $n$ and effective mass $m^*$.\cite{degiorgi,prl2005} Taking $\omega_c$= 5000 cm$^{-1}$, n=1.15$\times$10$^{21}$ cm$^{-1}$ and $m^*$=4.53 m$_e$ can be achieved.
Therefore, due to the violation of weak coupling BCS theory and the rather big effective mass, it is possible that Sr$_{3}$Ir$_{4}$Sn$_{13}$ is an example of such strong coupling CDW mechanism. Nevertheless, further studies are necessary on the unconventional CDW phase transition in such intermetallic compound.

In summary, we have successfully grown single crystal samples of Sr$_{3}$Ir$_{4}$Sn$_{13}$ and conducted careful characterizations by X-ray diffraction, electrical resistivity, magnetic susceptibility, and specific heat measurements. The (110)-plane optical measurements of Sr$_{3}$Ir$_{4}$Sn$_{13}$ single crystals were performed to elucidate the nature of the $T^*$ anomaly at 147 K. Bulk superconductivity is confirmed at 5 K. The optical spectroscopy reveals the formation of partial energy gap at the Fermi surface associated with structural phase transition. It is estimated that the opening of the gap leads to a 29 \% itinerant carriers loss and a substantial reduction of the carrier scattering rate. Unconventional CDW scenario is suggested to understand the origin of the second order phase transition. On the other band, a spectral weight transfer is seen from the low-energy mid-infrared region to the high-energy interband transition region above 5000 \cm both above and below the $T^*$ phase transition, which is attributed to the electron correlation effect.

~\\

We thank C. M. Varma for helpful discussions. This work is supported by the National Science Foundation of
China (11120101003, 11327806), and the 973 project of the
Ministry of Science and Technology of China (2011CB921701, 2012CB821403).

\bibliographystyle{apsrev4-1}

\end{document}